\documentclass[aps,pre,twocolumn,superscriptaddress]{revtex4}

\usepackage{graphicx}
\usepackage{color}

\begin{document}

\title{Bidirectional transport on a dynamic lattice}

\author{M. Ebbinghaus}

\email[]{ebbinghaus@lusi.uni-sb.de}

\affiliation{Laboratoire de Physique Th\'eorique, Univ. Paris-Sud, B\^at. 210, F-91405 Orsay Cedex, France}

\affiliation{CNRS, UMR 8627, F-91405 Orsay, France}

\affiliation{Fachrichtung Theoretische Physik, Universit\"at des Saarlandes, D-66123 Saarbr\"ucken, Germany}

\author{C. Appert-Rolland}

\email[]{cecile.appert-rolland@th.u-psud.fr}

\thanks{author to whom correspondence should be addressed}

\affiliation{Laboratoire de Physique Th\'eorique, UMR 8627, Universit\'e Paris-Sud XI, 91405 Orsay cedex, France}

\affiliation{CNRS, Orsay F-91405, France}

\author{L. Santen}

\email[]{l.santen@mx.uni-saarland.de}

\affiliation{Fachrichtung Theoretische Physik, Universit\"at des Saarlandes, D-66123 Saarbr\"ucken, Germany}

\date{\today}

\begin{abstract}
Bidirectional variants of stochastic many particle models for transport
by molecular motors show a strong tendency to form macroscopic clusters
on static lattices. Inspired by the fact that the microscopic tracks for molecular motors
are dynamical, we study the influence of different types of lattice dynamics on stochastic bidirectional transport. We observe a transition toward efficient transport (corresponding to the dissolution of large clusters) controlled by the lattice dynamics.
\end{abstract}


\maketitle

Microscopic models of stochastic transport like, e.g., the so-called asymmetric exclusion process (ASEP) or the zero range process (ZRP) have been extensively discussed in the past two decades \cite{blythe_e07}. The interest is partly due to the fact that these models play a key role in developing a general framework for statistical physics far from equilibrium~\cite{krug91}.
Moreover these models have been applied to many different transport problems, like road traffic \cite{chowdhury_s_s00}. More recently variants of the ASEP have been used in order to
describe intra-cellular transport phenomena driven  by molecular motors \cite{lipowsky_k_n01,parmeggiani_f_f03}, i.e., proteins that
are able to transport cargos along the filament network of biological cells~\cite{alberts02}.
In these models
particles can also attach to and detach from the filament, leading to a finite
path length of the molecular motors \cite{lipowsky_k_n01,parmeggiani_f_f03,tailleur_e_k09}.
Remarkably, 
if open boundary conditions are applied,
this extended model 
is able to predict the experimentally observed bulk localization of high and
 low density domains \cite{nishinari05}
 although many structural aspects of the filaments and
 motors have been neglected.

 In contrast to the success of these models describing the
 unidirectional collective motion of molecular motors
in motility assays, it is still an issue to understand
the relevant mechanisms involved in intracellular bidirectional
transport.
Several models of bidirectional transport have been suggested that consider positional exchange of particles with different moving directions on the same track (see, e.g., \cite{arndt_h_r98} and \cite{korniss_s_z99}). However, this family of models cannot be used to describe the motion of oppositely moving molecular motors which cannot permeate each other on the same track. This scenario is consistent 
with recent findings for certain members of the kinesin and dynein superfamily, moving in opposite directions on microtubules (MTs, which are polar filaments and a part of the cytoskeleton) and sharing the same binding site \cite{mizuno04}.

 Another difficulty arises from the fact that in real cells the
 volume surrounding the molecular tracks, i.e., the cytoplasm, is finite.
Confinement introduces a kind of memory effect for the motors that have detached
 from the filament,
 i.e., they are more likely to attach again in the
 vicinity of their detachment position.
 As a consequence, domains of high particle densities on the track and  in the diffusive surrounding reservoir co-localize \cite{ebbinghaus_s09,lipowsky_k_n01}.
 Both effects, i.e., inability of exchanges on the track and the explicit memory of the particle position, have been
 considered recently for static one- and multi-lane systems. It has been shown
\cite{ebbinghaus_s09}
 that for generic model parameters the formation of macroscopic clusters and therefore a
 breakdown of the flow on the filament is observed. Remarkably the formation of stable macroscopic  clusters does not depend on the particle density but rather on the number of particles in the system.
 Therefore the natural question arises: What are the
 minimal prerequisites in order to maintain bidirectional stochastic transport
 of interacting particles in small volumes
 when they cannot cross each other on the same track?
One suggestion has been made by Klumpp and Lipowsky \cite{klumpp_l04}
who considered that particles prefer to attach themselves in the neighborhood 
of motors of the same type, an effect that has been observed experimentally {\it in vitro}. 
For high densities and large differences
in the binding affinity this leads to a spontaneous formation of unidirectional
traffic.

In the present paper, we propose a completely different type of mechanism which could lead to efficient bidirectional transport on a single track through consideration of the filament dynamics. 
This extension of the model has been inspired by the experimentally
observed dynamics of the cytoskeleton.
Indeed, the MTs on which molecular motors move
are themselves highly 
dynamic~\cite{shemesh08}, due to nucleation, polymerization and
depolymerization, which occur on time
scales similar to those involved in motor
transport and are thus likely
to interfere with the motor dynamics.
Beyond the interest for
intracellular traffic, we shall give here a first example where a dynamically
driven jammed phase is hindered by the lattice dynamics.

\begin{figure}
 \includegraphics[scale=0.7]{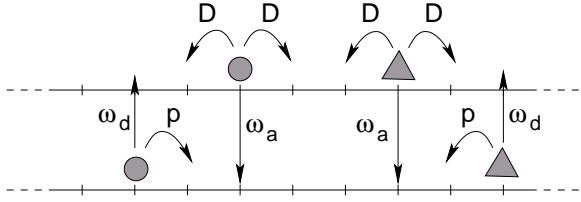}
  \caption{\label{fig:modeldef}Schematic representation of the particle dynamics (lattice dynamics is not included in this figure). Arrows indicate possible moves with corresponding rates which are symmetric for both particle species. We impose hard-core interaction on the lower lane (filament), while the particles on the upper lane are non-interacting. Periodic boundary conditions are considered.}
 \end{figure}

The model~\cite{ebbinghaus_s09} consists of two species of particles moving on two parallel lanes (Fig.~\ref{fig:modeldef}) with periodic boundary conditions. Along the lower lane, which mimics the polar filament, the particles perform directed motion (rate $p$) in the direction determined by the particle's species. In the upper lane, particles diffuse freely (rate $D$) and do not interact (sites on the upper lane can be multiply occupied). Attachment to the lower lane happens at rate $\omega_a$. The detachment rate $\omega_d$ is chosen to be smaller than the stepping rate $p$ in order to capture the processivity of molecular motors.
This setup is motivated by the quasi-one-dimensional geometry of axons although it largely (over-)simplifies the real structure. Please note that one can account for the geometry of the diffusive reservoir by tuning the attachment rate of the particles \cite{ebbinghaus_a_s10b}.
Here we consider only one processive lane. Counterintuitively, this simplification reduces, for a given particle density, the tendency to form macroscopic clusters as shown in \cite{ebbinghaus_s09}. 

As a new feature we add some lattice dynamics for the lower lane, i.e., some sites are eliminated and recreated. The diffusive upper lane remains unchanged. On the lower lane, particles can only occupy a site if this binding site exists. The attachment moves (rate $\omega_a$) are consequently rejected if the binding site has been eliminated. Additionally, a particle will automatically switch to the upper lane if it makes a forward step (rate $p$) onto an eliminated site or if the site which is occupied by the particle is eliminated.

In the following, we consider different types of lattice dynamics. In particular: (1.) a
simple realization of uncorrelated lattice dynamics. (2.) Dynamics which depends
on occupation by particles.
The results shown were obtained from Monte Carlo simulations over at
least $10^6$ steps with a constant set of parameters for the particle dynamics:
$p=1$, $\omega_d=0.02$, $\omega_a=0.33$ and $D=0.33$. The particle density has
been chosen high enough ($\rho_{tot}^\pm=0.5$) in order to observe large
clusters in the case of a static lattice even for small system sizes.

\begin{figure}
 \includegraphics[scale=0.3]{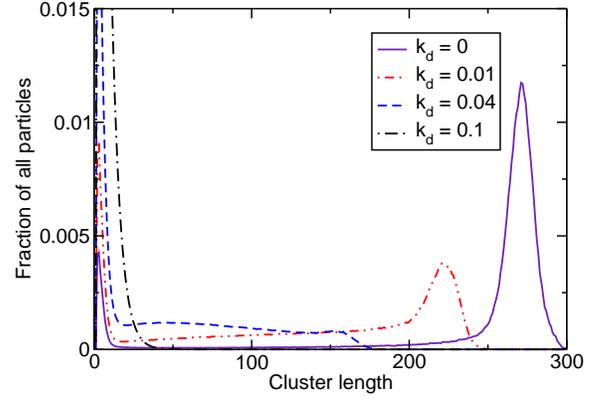}
  \caption{\label{fig:clusters}(color online). Distribution of cluster sizes in a system of size $L=1000$ with simple lattice dynamics (scenario 1). Recreation of lattice sites occurs at $k_p=1$. The black line corresponds to a static lattice.}
 \end{figure}

\begin{figure}
 \includegraphics[scale=0.3]{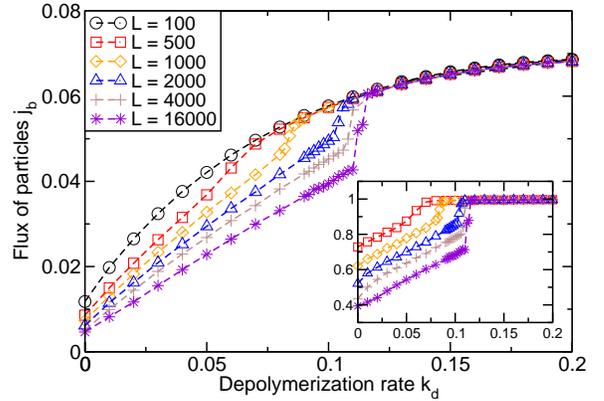}
  \caption{\label{fig:current}(color online). Flux of positive particles along the filament with simple lattice dynamics at $k_p=1$ for different system sizes $L$ and for the same density. If depolymerization is weak, the flux depends on the system size, but changes to a density-dependent state as large clusters disappear (see also Fig.~\ref{fig:clusters}). The inset shows the same data divided by the flux in the smallest system ($L=100$).}
 \end{figure}

First we consider some lattice dynamics independent
 of the configuration of particles:
a site of the lower lane is randomly eliminated at rate $k_d$ and recreated at rate $k_p$.
 This choice of the lattice dynamics
can be seen as a minimal model, with a limited number of parameters.
Network dynamics induces
an increase (resp. decrease) of the effective detachment rate $\omega_{d,\text{eff}}$ (respectively, attachment rate $\omega_{a,\text{eff}}$), which 
take the form
\begin{equation}
\label{eq:effrates}
\omega_{a,\text{eff}}=\omega_a\frac{k_p}{k_p+k_d};\quad\omega_{d,\text{eff}}=\omega_d+p\frac{k_d}{k_p+k_d}+k_d.
\end{equation}
As the depolymerization rate $k_d$ increases -- and accordingly the fraction of ``holes'' in the filament $k_d/(k_p+k_d)$ -- large clusters become less and less dominant until they disappear completely (Fig.~\ref{fig:clusters}).
This disappearance of large clusters is accompanied by a transition
from a size dependent to a size independent state (Fig.~\ref{fig:current}).
For large systems, flux vanishes below the transition, while it keeps
a finite value above the transition. As a result, the transition becomes
sharper for an increasing system size.

When the depolymerization rate $k_d$ is increased, first 
the flux along the filament (symmetric for both particle
species) increases too.
However, note that, if $k_d$ is too high, binding sites become increasingly sparse, and particles have fewer segments on which they can contribute to the total flux in the system. Hence, the flux of each particle species along the filament disappears for $k_d\to\infty$ and $k_p$ constant. This behavior entails the existence of an optimal value for $k_d$ 
(not seen on Fig.~\ref{fig:current}) at which the flux is maximized. In this
optimal regime with not too many holes in the filament,
blocking situations with at least two
particles of different species still occur frequently. As a consequence, the
maximum fluxes that we obtained are about one third of the flux in a
comparable single-species system~\cite{klumpp_l03}, which is nevertheless a
great improvement compared to the static filament case.

 \begin{figure}
 \includegraphics[scale=0.3]{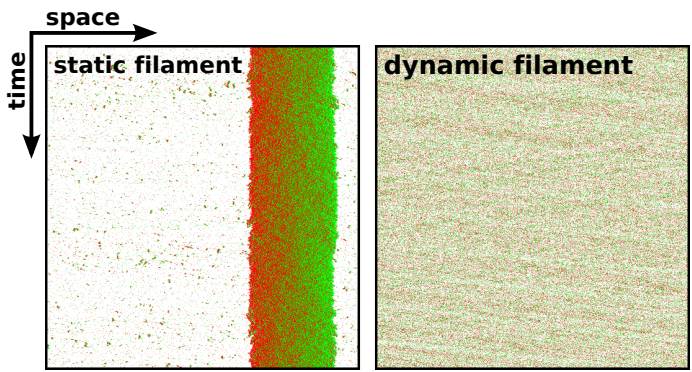}
  \caption{\label{fig:MFcurrent}(color online). Space-time plots of the filament occupation for a static (left) and a dynamic (right) filament in a system of size $L=1000$ and density $\rho_{tot}^\pm=0.5$. Red resp. green dots are particles with moving direction to the right resp. left, white spaces are unoccupied filament sites. The filament dynamics clearly induces a transition from a condensated to a homogenous phase. Note that considering a finite capacity of the diffusive lane would reinforce the jamming.}
\end{figure}

The degree of homogeneity can be estimated by comparison with a product-state.
Indeed, above the transition, particles are well-dispersed over the whole system (Fig.~\ref{fig:MFcurrent}) and are much less correlated than in the static filament case.
In a product-state approximation which neglects any correlations and if
translational invariance is assumed, the stationary state leads to~\cite{ebbinghaus_s09}
$\omega_{a,\text{eff}}\rho_u^\pm(1-\rho_b^\pm-\rho_b^\mp)=\omega_{d,\text{eff}}\rho_b^\pm$
where $\rho$ denotes the particle density, i.e., the number of particles divided by the system size. The indices $u$ and $b$ refer to the unbound state (upper lane) and bound state (lower lane) and $+/-$ signs denote the particle species moving either in positive or negative direction. In combination with the conservation of particles $\rho_{tot}^\pm=\rho_b^\pm+\rho_u^\pm$, the density of particles on the filament $\rho_b$ as well as the flux of one particle species along the filament $j_b^\pm=p\rho_b^\pm(1-\rho_b^\pm-\rho_b^\mp)$ can be calculated.

Below the transition, a big cluster is formed and the product-state approximation obviously fails.
Above the transition, the product-state prediction still overestimates (not shown) the flux.
This indicates that clusters made of a few particles
still play an important role. It is only for strong lattice dynamics
that the product-state solution regains validity.

\begin{figure}
 \includegraphics[scale=0.3]{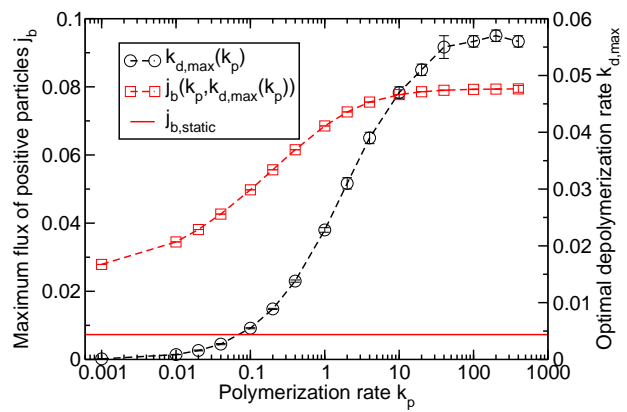}
  \caption{\label{fig:fluxmax}(color online). Maximum flux $j_b(k_p,k_{d,\max}(k_p))$ (red squares) and optimal depolymerization rate $k_{d,\max}(k_p)$ (black circles) at given polymerization rates $k_p$ in a system of size $L=1000$. For comparison, the flux in a system with a static lattice is drawn as a solid red line. Note that the horizontal axis is in logarithmic scale and, therefore, both quantities seem to saturate at high polymerization rates.}
 \end{figure}

In Fig.~\ref{fig:fluxmax}, simulation results for the optimal depolymerization rate $k_{d,\max}(k_p)$ maximizing the flux are drawn as a function of the polymerization rate. The maximum flux $j_b(k_p,k_{d,\max}(k_p))$ is shown as a function of $k_p$. One can see that the gain in transport capacity is obtained for a wide range of values of the polymerization rate $k_p$ -- an observation which supports the general validity of our results. Furthermore, the fraction of eliminated filament sites $k_d/(k_p+k_d)$ in the optimal regime decreases with increasing polymerization rate $k_p$ as the optimal value $k_{d,\max}$ saturates. This indicates that the optimal flux depends on the time interval a site persists rather than on the delay after which it is recreated.

In the second realization of lattice dynamics considered, a site is eliminated with rate $k_d$ {\em only} if that site is occupied.
This can be considered as a prototypical example of a coupling
between MT dynamics and transport by molecular motors.
On average, the fraction of existing sites on the lower lane becomes $k_p/\left[k_p+(\rho_b^++\rho_b^-)k_d\right]$ and Eq.~(\ref{eq:effrates}) is modified accordingly. 

It turns out that the results are qualitatively the same as obtained for the first type of dynamics. The main difference is that  the flux in the homogenous phase is higher than in the first scenario. In this phase, when a motor is moving freely, it encounters less holes than in the case of scenario 1, as empty sites cannot be eliminated anymore. Processive runs along the filament are thus less frequently interrupted.
 By contrast, in the condensated phase, there is almost no difference to the first scenario as the transport capacity is limited by the large clusters, where  all filament sites are occupied by particles anyway.
The transition to a density-dependent efficient state is maintained
if one generalizes scenario 2 by requiring that a higher number of particles has to accumulate in order to put enough strain on the filament to break.

Finally we mention briefly another kind of dynamics. Biopolymers, such as actin filaments or MTs, show a characteristic type of dynamics under certain conditions, termed tread-milling~\cite{alberts02}, for which
both ends of the filament move with the same average velocity.
We have checked that, for a fairly simplified model consisting of regularly
spaced holes in the lower lane which propagate synchronously but stochastically
through the system, the aforementioned transition toward efficient
transport is still observed.
Note that, since the holes move only in one direction, the two species
of particles are affected differently. The flux of the two species therefore
depends on the moving direction along the filament. For both species we remark
a considerable increase of the flux and reach a maximum current comparable to
the one in the previous scenarios.

To summarize, we have presented a model for bidirectional
transport on a one-dimensional dynamic lattice
coupled to a confined diffusion reservoir.
While on a static lattice persistent
clusters form which inhibit efficient
transport~\cite{ebbinghaus_s09}, we 
have found the counterintuitive effect that the
transient suppression of filament sites dramatically {\em enhances}
the transport capacity.
Indeed, the inhibition of large
clusters leads to a transition toward a homogenous
state characterized by efficient
transport. This is a new mechanism in the phenomenology of
dynamic phase transitions.
This transition separates a size-dependent (jammed) state from
a density dependent one.
It is robust in the sense that it is obtained for quite
different types of lattice dynamics.
The actual transport capacity of the system rather depends on the optimal lifetime of a binding site than on the details of the filament dynamics. The lifetime has to be short enough in order to avoid jam formation and long enough in order to direct the transport.

Some insight could be gained from an analysis similar
to the one found in \cite{nagar_m_b06}
for the symmetric motion of non-interacting particles
in fluctuating energy landscapes. However, note that here the effective ``potential''
landscape emerges spontaneously from the particle jamming.

Finally we would also like to discuss the relevance of our model results for axonal transport.
In view of our results for bidirectional transport on static lattices,
which generically leads to jamming,
it is rather surprising, but of course necessary, that transport in
these real one-dimensional axonal structures would be at all stable
and efficient.
This is all the more surprising given that the model
overestimates the diffusivity of the detached particles, which
should be reduced in real systems due to the size and interactions
of the vesicles in the cytoplasm. These results give strong evidence that additional
effects must come into play in order to stabilize motor driven transport in axons and reduce the
tendency to form large particle clusters. The mechanism suggested in this work is based on the
fact that the filament dynamics limits the size of particle clusters.
A more detailed modeling of motor driven transport
in axons is difficult since the experimental findings
concerning the motor interactions and the MT-dynamics are so far incomplete and subject to interpretation:
while it is well established that the plus-ends of
MTs undergo polymerization events toward the synapse, the pathlength
of the growing plus-ends as well as the dynamics of
the minus ends is not known yet.
Besides the robustness of our results
with respect to the lattice dynamics details,
the importance of the
lattice dynamics is supported by experimental results on transport in axons, where the strong
correlation between MT dynamics and vesicle transport has been demonstrated \cite{shemesh_s10}. In this context, we have observed
that lane formation, which could be invoked as an alternative
mechanism for bidirectional transport \cite{klumpp_l04}, is impeded by 
the dynamics of MTs.
Thus it seems that it is necessary to take into account the lattice dynamics
in order to gain understanding in axonal transport.
Although our picture of  bidirectional transport is
qualitatively consistent with the experimental findings it is still an issue to develop a more
quantitive description of real axonal transport. A fully validated model could then be used in
order to investigate mechanisms that are underlying clinically relevant transport defects in nerve
cells~\cite{stokin05,shemesh08}.

\begin{acknowledgments}

ME would like to thank the DFG Research Training Group GRK 1276 for financial support.

\end{acknowledgments}


\end{document}